\documentclass[11pt]{article}

\usepackage[utf8]{inputenc}
\usepackage[T1]{fontenc}
\usepackage{amsmath,amssymb}
\usepackage{graphicx}
\usepackage[margin=1in]{geometry}
\usepackage{hyperref}
\hypersetup{pdftitle={Implementing the Spec Growth Engine},
            pdfauthor={Hartwig Grabowski}}
\usepackage[numbers,sort&compress]{natbib}
\usepackage{microtype}
\usepackage{xcolor}
\usepackage{booktabs}
\usepackage{tabularx}
\usepackage{enumitem}
\usepackage{listings}
\usepackage{upquote}
\usepackage{tikz}
\usetikzlibrary{arrows.meta,positioning,shapes.geometric,fit,backgrounds,calc,decorations.pathreplacing}
\usepackage{caption}
\usepackage{subcaption}
\usepackage{mdframed}

\definecolor{teal}{HTML}{1C7293}
\definecolor{teallight}{HTML}{E1F0F2}
\definecolor{coral}{HTML}{B85042}
\definecolor{corallight}{HTML}{F6E7E3}
\definecolor{green}{HTML}{3B6D11}
\definecolor{greenlight}{HTML}{EAF3DE}
\definecolor{sand}{HTML}{F1EFE8}
\definecolor{orange}{HTML}{BA7517}
\definecolor{orangelight}{HTML}{FAEEDA}
\definecolor{darktext}{HTML}{21303A}
\definecolor{mutedtext}{HTML}{5F5E5A}

\newmdenv[
  backgroundcolor=corallight,
  linecolor=coral,
  linewidth=1.2pt,
  roundcorner=4pt,
  innertopmargin=6pt,
  innerbottommargin=6pt,
]{problembox}

\newmdenv[
  backgroundcolor=teallight,
  linecolor=teal,
  linewidth=1.2pt,
  roundcorner=4pt,
  innertopmargin=6pt,
  innerbottommargin=6pt,
]{conceptbox}

\newmdenv[
  backgroundcolor=greenlight,
  linecolor=green,
  linewidth=1.2pt,
  roundcorner=4pt,
  innertopmargin=6pt,
  innerbottommargin=6pt,
]{solutionbox}

\newcommand{\Hp}{H$'$}

\title{\textbf{Implementing the Spec Growth Engine}\\
  \large Preventing Spec-Code Divergence, and Growing the Spec with Agents}

\author{
  Hartwig Grabowski\\
  \small Hochschule Offenburg\\
  \small \texttt{hartwig.grabowski@hs-offenburg.de}\\
  \small ORCID: \href{https://orcid.org/0009-0001-4300-2626}{0009-0001-4300-2626}
}

\date{\today}

\begin{document}
\maketitle

\begin{abstract}
The Spec Growth Engine anchors AI-assisted software development in a graph of
specifications that the code is coupled to~\cite{grabowski2026sge}.  This paper
describes its implementation, which serves two tasks and keeps them apart as two
layers.  The first layer \emph{prevents spec-code divergence}: a deterministic
engine validates the spec graph, compares it with the code's import graph, earns
a node's \emph{verified} status from recorded test evidence, and classifies
every change by what it can break -- without calling a model.  The second layer
\emph{grows the spec with agents}: an intent author, a planner and a coder, each
played by its own model, extend the graph in rounds, and a deterministic rule
decides after each round whether the run goes on.  How much of the human's
judgement is delegated is set by three independent switches -- a draft gate, a
delegation for breaking changes, and the run mode -- which, with two ways of
laying a project's floor, give eighteen ways to run a project.  We describe
each of them, the gates, requests and waivers through which agents and the human
communicate, and the spectrum of operation from entirely manual work to an
unsupervised run whose decisions the human reviews afterwards.  Throughout, one
claim holds the design together: an autonomous run is worth only as much as the
deterministic instance that measures it.
\end{abstract}

\section{Introduction}
\label{sec:intro}

The companion paper~\cite{grabowski2026sge} introduced the Spec Growth Engine as
a method: software grows as a graph of \texttt{SPEC.md} files, one per node, each
stating a contract; the code is coupled to that graph; and divergence between the
two is a blocking condition rather than a slow erosion.  It argued the design
from its failure modes -- context explosion and silent spec-code drift -- and
from established foundations~\cite{parnas1972,murphy1995,ford2022,freeman2009}.

This paper describes the implementation: an engine, a command-line interface and
a graphical \emph{cockpit}, developed and tested in a series of agent runs on
small, concrete tasks.  Building it forced a distinction that the method had left
implicit.  The implementation serves two tasks of a different nature:

\begin{enumerate}[leftmargin=*]
  \item \textbf{Preventing spec-code divergence.}  Given a graph and a code base,
        say -- deterministically, and on every commit -- whether they agree:
        whether every import the code makes is a declared dependency, whether
        every file belongs to a node, whether a node claimed as done has the test
        evidence to be, and whether a change breaks a promise others rely on.
  \item \textbf{Growing the spec with agents.}  Given a goal, let agents write
        the node intents, the specifications and the code -- and decide, after
        each round of their work, whether the goal is met, whether they go on,
        or whether they are waiting for a human.
\end{enumerate}

The first task needs no model and must not use one: its value is that it is
reproducible, and that it blocks.  The second task consists of judgement and
cannot do without models.  The implementation therefore keeps them apart as two
\emph{layers} (Section~\ref{sec:layers}).  Layer~1 is a complete use of the
engine on its own -- a person, a text editor and the gate.  Layer~2 rests on
layer~1 as its floor and its witness:

\begin{conceptbox}
\textbf{The claim the design rests on.}  An autonomous run is worth only as much
as the deterministic instance that measures it afterwards.  An agent that says
\emph{done} proves nothing; a gate that blocks, a test that is recorded, and a
change set that says what moved are what make its claim checkable.  Layer~1
does not need layer~2; layer~2 cannot exist without layer~1.
\end{conceptbox}

\paragraph{Contributions.}
\begin{itemize}[leftmargin=*]
  \item A deterministic floor for layer~1: structural validation, drift
        detection against the import graph, an earned status lifecycle, a change
        set that classifies every change by its gate, and an \emph{accepted
        state} that lets the gate measure changes nobody saved by hand
        (Sections~\ref{sec:floor} and~\ref{sec:gates}).
  \item Two channels from agents to the human that grant nothing by themselves --
        \emph{decision requests} and \emph{waivers} (Section~\ref{sec:requests}).
  \item Three independent autonomy switches instead of a single ladder, and the
        eighteen combinations they form with two ways of laying a project's floor
        (Sections~\ref{sec:switches} and~\ref{sec:eighteen}).
  \item A run driver that lets agents work in rounds under a deterministic rule,
        including an unsupervised mode in which a reviewer model decides in the
        human's place and the human reviews every decision afterwards
        (Section~\ref{sec:operation}).
\end{itemize}

Section~\ref{sec:engine} describes the engine, Section~\ref{sec:operation} how it
is operated -- from entirely by hand to unsupervised -- and
Section~\ref{sec:example} follows one project from an empty directory to a met goal,
and gives the same goal to a supervised and an unsupervised agent run.

\section{The Engine}
\label{sec:engine}

\subsection{Two Layers, Six Roles}
\label{sec:layers}

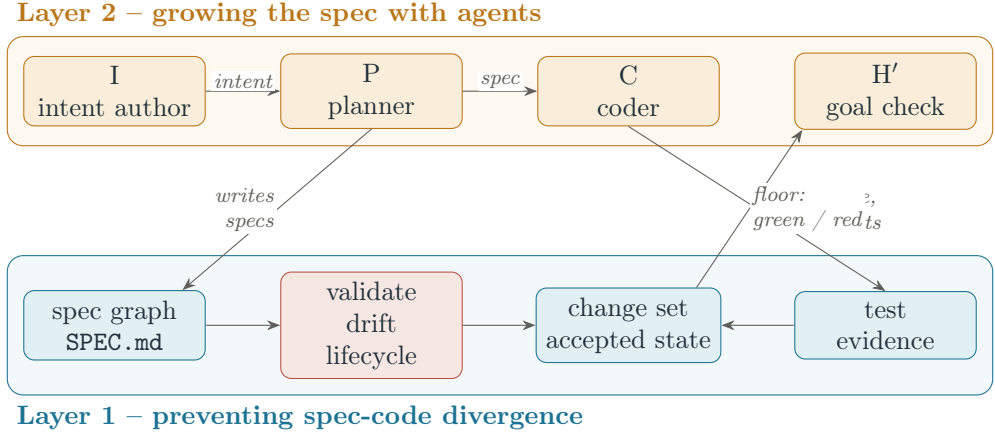
\begin{figure}[h]
\centering
\begin{tikzpicture}[
  every node/.style={font=\small},
  layer/.style={draw, rounded corners=6pt, inner sep=6pt},
  comp/.style={draw, rounded corners=4pt, minimum width=2.4cm, minimum height=0.8cm,
               fill=teallight, draw=teal, text=darktext, align=center},
  agent/.style={draw, rounded corners=4pt, minimum width=2.4cm, minimum height=0.8cm,
                fill=orangelight, draw=orange, text=darktext, align=center},
  gate/.style={draw, rounded corners=4pt, minimum width=2.4cm, minimum height=0.8cm,
               fill=corallight, draw=coral, text=darktext, align=center},
  arrow/.style={-{Stealth[length=5pt]}, draw=mutedtext},
  lbl/.style={font=\scriptsize\itshape, text=mutedtext, fill=white, inner sep=1pt},
]
  \node[agent] (i) at (0, 3.1)    {I\\intent author};
  \node[agent] (p) at (3.4, 3.1)  {P\\planner};
  \node[agent] (c) at (6.8, 3.1)  {C\\coder};
  \node[agent] (h) at (10.2, 3.1) {\Hp\\goal check};
  \draw[arrow] (i) -- node[above,lbl]{intent} (p);
  \draw[arrow] (p) -- node[above,lbl]{spec} (c);

  \node[comp] (g) at (0, 0)    {spec graph\\\texttt{SPEC.md}};
  \node[gate] (v) at (3.4, 0)  {validate\\drift\\lifecycle};
  \node[comp] (s) at (6.8, 0)  {change set\\accepted state};
  \node[comp] (e) at (10.2, 0) {test\\evidence};

  \draw[arrow] (g) -- (v);
  \draw[arrow] (v) -- (s);
  \draw[arrow] (e) -- (s);

  \draw[arrow] (p.south) -- node[left,lbl,align=right]{writes\\specs} ($(g.north)+(0.9,0)$);
  \draw[arrow] (c.south) -- node[right,lbl,align=left]{writes code,\\records tests} (e.north);
  \draw[arrow] ($(s.north)+(0.9,0)$) -- node[right,lbl,align=left]{floor:\\green / red} ($(h.south)+(-1.1,0)$);

  \begin{scope}[on background layer]
    \node[layer, draw=orange, fill=orangelight!40, fit=(i)(p)(c)(h),
          label={[font=\small\bfseries, text=orange, anchor=south west]north west:Layer 2 -- growing the spec with agents}] {};
    \node[layer, draw=teal, fill=teallight!40, fit=(g)(v)(s)(e),
          label={[font=\small\bfseries, text=teal, anchor=north west]south west:Layer 1 -- preventing spec-code divergence}] {};
  \end{scope}
\end{tikzpicture}
\caption{The two layers.  In layer~2, agents write intents, specifications and
  code.  Layer~1 measures what they wrote -- structure, drift, status, the
  gate of every change -- and the result decides, with the goal check, whether a
  run goes on.  Layer~1 works without layer~2 and without any model.}
\label{fig:layers}
\end{figure}

Figure~\ref{fig:layers} shows the two layers and what passes between them.
Layer~1 consists of files -- \texttt{SPEC.md} per node, \texttt{ARCHITECTURE.md}
for what holds everywhere, and a \texttt{.sge/} directory for the engine's records
-- and of a gate that reads them.  Layer~2 consists of a goal (\texttt{TASK.md}),
the manuals the agents read, and the runs.

Six roles operate the engine (Table~\ref{tab:roles}).  Four of them call a model;
the engine calls none, and the human is always a person.

\begin{table}[h]
\centering
\caption{The roles and who plays them.}
\label{tab:roles}
\small
\begin{tabular}{@{}clp{6.3cm}p{3.6cm}@{}}
\toprule
& \textbf{Role} & \textbf{Does} & \textbf{Played by} \\
\midrule
\textbf{H}  & Human architect & The goal, the boundaries, public contracts,
                                root invariants; approves consequential changes;
                                grants waivers. & A person, always. \\
\textbf{I}  & Intent author   & Says what each node is for -- and how the graph
                                grows to meet the goal.  Writes no spec and no code. &
                                H in the cockpit, or an agent in a run. \\
\textbf{P}  & Planner         & Turns an intent into a \texttt{SPEC.md}: contract,
                                acceptance criteria, design. & A model (chat call). \\
\textbf{C}  & Coder           & Implements one node from its specification. & An agent. \\
\textbf{\Hp} & Reviewer       & Judges what the engine cannot: drafts, breaking
                                changes, whether a run met its goal. & A model, never
                                a person. \\
\textbf{E}  & Engine          & Derives, validates, blocks, decides the rounds of a
                                run.  Authors no intent. & Code; calls no model. \\
\bottomrule
\end{tabular}
\end{table}

The separation of I, P and C is what makes the models separable: each model-backed
role has its own endpoint, so a run can be played by one model for everything or
by a different one per role.  Nobody marks their own work: C implements what P
specified from what I intended, and \Hp{} -- a separate call that does not see
what the author said about its work -- judges the result.

\subsection{The Deterministic Floor}
\label{sec:floor}

Every architectural entity is a node with exactly one \texttt{SPEC.md}, whose
\texttt{kind} field names its level: \texttt{system} (the root),
\texttt{container} or \texttt{component}.  Specifications end at the component,
which nests as deep as it needs to; the code is not a node but what a
component's specification generates and is measured against, and on one
ownership line the most specific owner of a file wins.  A node's front matter
carries the contract (\texttt{provides}, invariants, error behaviour), the
acceptance criteria, the declared dependencies (\texttt{depends\_on}), the code
it owns (\texttt{code} globs) and its status.  Three checks read the graph and
the code together; together they are the \emph{floor}, and the gate blocks when
any of them fails:

\begin{description}[leftmargin=1.2cm, style=nextline]
  \item[validate] The graph is well formed: every parent exists, every declared
        dependency names a node with a contract and lies outside the node's own
        ownership line, the specifications parse, requests and waivers are well
        formed (Section~\ref{sec:requests}).
  \item[drift] The code agrees with the graph: the engine builds an evidence
        graph from the imports the code actually makes -- for Python,
        TypeScript/JavaScript, Java/Kotlin and Dart -- and a cross-node import
        that is not a declared dependency is a blocking error.  A file no node
        owns, and a code glob that matches nothing, are reported as well.
  \item[lifecycle] A status is earned.  A node moves from \texttt{specd} through
        \texttt{implemented} to \texttt{verified}; \texttt{verified} requires
        recorded, passing test evidence for its acceptance criteria and no open
        waiver, and no node may be \texttt{verified} while a child is not.
\end{description}

Test evidence is recorded by the engine when it runs the project's tests at its
build roots (\texttt{cli.py test}), stamped with the run's start, so that a
later change voids it.  The floor runs offline and deterministically, fast
enough to be a git pre-commit hook: the implementation installs one, so that a
drifting commit is refused.

\subsection{Gates: SOFT and HARD}
\label{sec:gates}

The floor says whether graph and code agree.  It does not say whether a
\emph{change} to the graph is harmless.  For that, the engine computes a
\emph{change set} for every changed specification: which \emph{gate} the change
falls under, and whom it reaches.  Table~\ref{tab:gates} gives the rules; the
strictest reason decides.

\begin{table}[h]
\centering
\caption{Gate levels of a change, as the engine classifies them.}
\label{tab:gates}
\small
\begin{tabularx}{\textwidth}{@{}lX@{}}
\toprule
\textbf{Gate} & \textbf{Changes that fall under it} \\
\midrule
\textbf{HARD} & A new root or container; an id, kind or parent changed; the root's
  contract changed; a \emph{breaking} contract change (an operation removed or
  reworded, an invariant or the error behaviour changed) when a node depends on
  it, the node is a container or the root, or it is declared protected; a node
  deleted that others depend on; removing a protection; granting a waiver; any
  change to \texttt{ARCHITECTURE.md}. \\
\textbf{SOFT} & A new component; operations only \emph{added} to a contract; a
  breaking change nobody can be broken by yet; acceptance criteria and code
  globs changed; declaring a protection. \\
\textbf{AUTO} & \texttt{depends\_on} or status changed -- the drift and lifecycle
  checks decide. \\
\textbf{NONE} & Design, prose, a request. \\
\bottomrule
\end{tabularx}
\end{table}

``Breaking'' is read off the text and fails safe: whether an invariant was
tightened or widened cannot be read off a sentence, so any change to one counts.
The only additive change the engine can prove is a pure addition to
\texttt{provides}.  Alongside the gate, the change set lists the consumers of a
changed contract, which of them mention what was removed, which evidence was
recorded against the old contract -- and, entry by entry, what changed, before
and after.  A consumer is a node that depends on the contract, or a parent whose
code imports it: a parent uses its sub-component along the ownership line, where
no declared dependency may name it.  The entries are the counterpart of failing
safe: the gate says only
\emph{that} an invariant changed, and the human asked to clear the change has to
read \emph{which}.  It is shown whatever the gate, each part under its own --
the change takes the highest -- so that the part that makes a change HARD
stands apart from the SOFT and AUTO parts that came with it; only design and
prose, which gate nothing, stay out.

\paragraph{The accepted state.}  In the cockpit, a change is measured against
the file on disk at the moment of saving, and a HARD change asks the human
before it is saved.  An agent writing files has no such moment.  The engine
therefore keeps an \emph{accepted state} -- every node's specification as it
was last accepted, and \texttt{ARCHITECTURE.md} -- and measures against it.  On a
passing gate, the engine itself accepts every change at SOFT or below -- and says
which: run from the cockpit, the gate's report shows each change it accepted,
with its diff, so that what was accepted without the human is still there for
the human to read.  A HARD
change is not blocked: it stays \emph{pending}, and the changed node and the
consumers of a broken contract cannot reach \texttt{verified} until the human
accepts it.  Rejecting it, with a reason, turns it into a blocking error that
carries the reason to the agent until the node is restored or changed
differently.  A node or a subtree can be declared \texttt{protected}, which
makes every breaking change to it HARD even when nothing in the graph depends on
it -- for boundaries whose consumers sit outside the graph, such as a command
line, an API or a file format.

\subsection{Requests and Waivers}
\label{sec:requests}

Some decisions are not the asker's to make, and some rules cannot be met yet.
In both cases the engine gives the asker a way to \emph{record} the situation
that grants nothing by itself -- because the alternative, observed in an early
run, is that the agent deviates anyway and describes it in prose that no gate
reads.

\paragraph{Decision requests.}  A node that needs a decision above its authority
records it on itself: the target -- \texttt{ARCHITECTURE.md} for an invariant, or
a node the asker may not grow, such as a container -- the proposal, worded as it
should appear there, and the reason.  The request grants
nothing; it holds the node back from \texttt{verified}, it is reported on every
run, and the human resolves it by changing the target -- or rejects it,
optionally recorded as an architecture decision (ADR~\cite{nygard2011}) so that
it is not asked again.  Between two nodes the asker grows itself there is no
request: that is a revision of its own node.

\paragraph{Waivers.}  A waiver is a typed, owned and time-boxed permission to
deviate from a rule.  Requesting and granting are separate acts by separate
parties: anyone, an agent included, may write a request with a reason; only the
human grants, with an owner and an expiry.  A waiver has one of four kinds --
\texttt{pending-external-evidence}, \texttt{known-gap}, \texttt{deferred-test}
and \texttt{frontier-stub}, the last of which marks a deliberate fake at the
edge of a slice and keeps its node at \texttt{specd}.  Past its expiry, a waiver
blocks the gate, so a deviation must be renewed deliberately or resolved.

\paragraph{The human's inbox.}  Requested waivers, decision requests and pending
HARD changes arrive in one place -- the cockpit's \emph{Requests} window, or
\texttt{cli.py pending} -- with the actions that resolve them.  For a pending
HARD change, \Hp{} can prepare a \emph{brief}: what changes, whom it reaches,
the risks, non-breaking alternatives and a recommendation.  The brief is bound
to that exact change and goes stale when the node changes again.

\subsection{Three Switches}
\label{sec:switches}

How much of the human's judgement is delegated was first conceived as a single
ladder of autonomy levels.  The implementation showed that it is not one scale:
three different things are delegated, and each has a switch of its own.

\begin{description}[leftmargin=1.2cm, style=nextline]
  \item[The draft gate: D0, D1, D2.]  Who looks at a drafted \texttt{SPEC.md}
        before it is saved.  At \textbf{D0} the human alone.  At \textbf{D1}
        \Hp{} reviews the draft first and lists what it finds -- a contract that
        says \emph{how} instead of \emph{what}, a criterion no test could check,
        a capability that already exists elsewhere -- and the human still saves.
        At \textbf{D2} \Hp{} may save the draft itself, but only a SOFT one it
        is sure of (confidence at least~0.6); if it wants changes, the planner
        revises the draft at most twice first, and everything else -- a HARD
        change, a protected node, a draft \Hp{} rejects or is unsure about --
        comes to the human.  The gate stops at D2, because a draft is not on
        disk and cannot be migrated.
  \item[The delegation: on or off.]  Whether \Hp{} may accept a \emph{pending}
        HARD change itself.  Not a grade above D2 but an answer to one question,
        set per project, by the human only.  Its scope is narrow: a breaking
        change to a component contract that nodes in the graph consume --
        never a protected zone, a container, the root, \texttt{ARCHITECTURE.md},
        a new node or a deletion.  It applies only once the engine finds the
        change \emph{migrated} -- the node and every consumer tested after the
        change, no consumer still naming a removed operation -- and \Hp{}
        recommends it.  Every acceptance by \Hp{} waits for the human to confirm
        or overrule it, and overruling restores the previous accepted
        specification.
  \item[The run: off, on or unsupervised.]  Who authors node intent.
        \textbf{Off}: the human, in the cockpit.  \textbf{On}: an agent, which
        derives node intents from the goal and has P and C do the rest; what is
        above a node waits for the human.  \textbf{Unsupervised}: as on, and
        \Hp{} answers the questions above a node between rounds, in the
        human's place (Section~\ref{sec:unsupervised}).
\end{description}

\noindent Whatever the switches say, the floor always runs and always blocks:
\Hp{} judges what the engine cannot, and never overrides it.

\subsection{Eighteen Ways to Run a Project}
\label{sec:eighteen}

A project starts with its \emph{floor} laid in one of two ways.  \textbf{Guided}:
the cockpit's \emph{New Project} dialog lays a root, a first container with its
code layout and build manifest (from an archetype: Python with uv, Java with
Maven, TypeScript with npm, Rust with Cargo), the invariants that hold everywhere
-- four defaults and four questions, on persistence, concurrency, network and
anything else -- the accepted state, the two agent manuals and, if an agent is to
build it, the goal.  \textbf{Bare}: a root and a thin \texttt{ARCHITECTURE.md}
that the human fills in.  The floor does not change how a switch behaves; it
changes what exists when the work starts.

Combined with the switches, the floor gives eighteen ways to run a project
(Table~\ref{tab:eighteen}).  The draft gate matters only where a human drafts:
a run drafts through the command line, which no slider reaches, so its rows
carry a dash.  The run in unsupervised mode requires the delegation, so it has
no row with the delegation off.  The table's decisive column is the fifth: when
the human is asked -- \emph{before} the tool acts, or \emph{after} it has.

\begin{table}[p]
\centering
\caption{Eighteen ways to run a project: two floors, the draft gate, the
  delegation and the run -- and when each one asks the human.}
\label{tab:eighteen}
\footnotesize
\renewcommand{\arraystretch}{1.15}
\begin{tabularx}{\textwidth}{@{}llll>{\raggedright\arraybackslash}X>{\raggedright\arraybackslash}p{2.8cm}@{}}
\toprule
\textbf{Floor} & \textbf{Draft} & \textbf{Deleg.} & \textbf{Run} &
  \textbf{You are asked \ldots} & \textbf{It needs} \\
\midrule
guided & D0 & off & off & BEFORE every save: you press it. Every HARD change
  waits for you. & P, for drafting \\
guided & D0 & on  & off & As with the delegation off: every HARD change asks
  you when it is saved, so \Hp{} has nothing left to accept. & P, for drafting \\
guided & D1 & off & off & BEFORE every save, with what \Hp{} found in front of
  it. Every HARD change waits for you. & P, \Hp{} \\
guided & D1 & on  & off & As with the delegation off. & P, \Hp{} \\
guided & D2 & off & off & AFTER a SOFT draft saved itself (the log, the change
  set) -- and BEFORE a HARD one, which escalates and waits for you. & P, \Hp{} \\
guided & D2 & on  & off & As with the delegation off. & P, \Hp{} \\
guided & --  & off & on  & AFTER each round: its decision requests and its HARD
  changes, in Requests. NEVER before it acts -- a run is unattended by
  definition. & I, P, C, \Hp{} \\
guided & --  & on  & on  & As above, and AFTER \Hp{} accepted a migrated breaking
  change: confirm or overrule. Still never before it acts. & I, P, C, \Hp{} \\
guided & --  & on  & unsup. & AFTER the run, never during it: what \Hp{} decided
  in your place -- the requests it answered, the HARD changes it accepted -- to
  keep or take back. Yours still: the goal, the zones you declared. &
  I, P, C, \Hp{} \\
\midrule
bare & D0 & off & off & as above & P; the invariants are yours \\
bare & D0 & on  & off & as above & P; the invariants \\
bare & D1 & off & off & as above & P, \Hp{}; the invariants \\
bare & D1 & on  & off & as above & P, \Hp{}; the invariants \\
bare & D2 & off & off & as above & P, \Hp{}; the invariants \\
bare & D2 & on  & off & as above & P, \Hp{}; the invariants \\
bare & --  & off & on  & as above & all four; the invariants, the goal, the
  manuals \\
bare & --  & on  & on  & as above & as above \\
bare & --  & on  & unsup. & as above & as above \\
\bottomrule
\end{tabularx}
\end{table}

The columns, in turn:

\begin{description}[leftmargin=1.2cm, style=nextline]
  \item[Floor.]  How the project was laid: guided or bare.  A bare floor needs
        the invariants written by hand before anything measured against them
        means much -- and a run on a bare floor needs the goal and the manuals
        that the guided floor writes.
  \item[Draft gate.]  Who looks at a draft before it is saved: D0, D1 or D2
        (Section~\ref{sec:switches}).  It applies to drafts P wrote; a
        specification written by hand is saved like D0.
  \item[Delegation.]  Whether \Hp{} may accept a pending, migrated breaking
        change.  Without a run it changes nothing: in the cockpit a HARD change
        asks the human when it is saved, so nothing is left pending.  It matters
        when someone else changes the specifications -- an agent.
  \item[Run.]  Whether an agent authors the intents, and whether \Hp{} answers
        for the human while it runs.
  \item[You are asked.]  When the human is involved -- the column that differs
        day to day.  The manual rows ask before; the run rows only after.
  \item[It needs.]  Which roles need a model endpoint.  What the column names is
        what the \emph{help} costs; the engine itself needs no key.  The
        nineteenth way needs nothing at all: specifications written by hand,
        with the gate, drift, evidence and the accepted state all working.
\end{description}

\paragraph{Who clears a change, and where.}  It is tempting to read the switches
as one ladder: if \Hp{} may accept a HARD change, surely it may accept a SOFT
one, and the delegation should imply D2.  It does not, because the two switches
act on different changes in different places (Table~\ref{tab:clears}).  In a
run, SOFT changes are cleared anyway -- by the engine, not by D2; in the cockpit
the delegation has nothing to act on; and the draft gate plays no part in a run,
supervised or not, since the agent drafts through the command line, which no
slider reaches.  What \Hp{} judges differs as well: at D1 and D2 the
\emph{quality} of a draft -- what rather than how, testable criteria, reuse --
under the delegation whether a broken promise has been \emph{migrated}.

\begin{table}[h]
\centering
\caption{Who clears a change, and where.}
\label{tab:clears}
\small
\begin{tabularx}{\textwidth}{@{}l>{\raggedright\arraybackslash}X>{\raggedright\arraybackslash}X@{}}
\toprule
& \textbf{In the cockpit (the human drafts)} & \textbf{In a run (an agent writes)} \\
\midrule
\textbf{SOFT} & The draft gate: D0 the human, D1 the human after \Hp{} reviewed it,
  D2 \Hp{}. & The engine, on a passing gate -- wherever the draft gate stands. \\
\textbf{HARD} & The human, always, when saving. & The human -- or, with the delegation
  on, \Hp{} for one narrow kind: a breaking change to a consumed component contract
  that the engine found migrated. \\
\bottomrule
\end{tabularx}
\end{table}

\section{Operation: From by Hand to Unsupervised}
\label{sec:operation}

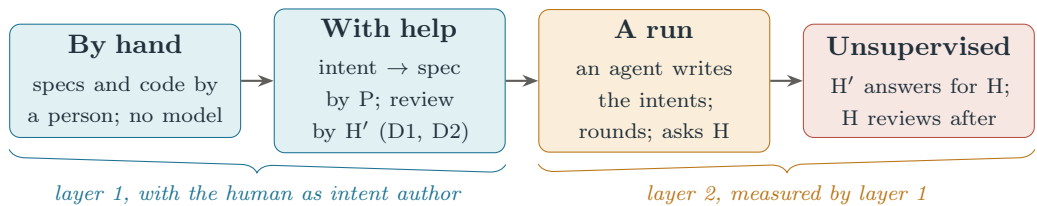
\begin{figure}[h]
\centering
\begin{tikzpicture}[
  every node/.style={font=\small},
  stage/.style={draw, rounded corners=4pt, minimum width=3.0cm, minimum height=1.5cm,
               text width=2.8cm, align=center, text=darktext},
  arrow/.style={-{Stealth[length=6pt]}, draw=mutedtext, line width=0.8pt},
]
  \node[stage, fill=teallight, draw=teal]      (a) at (0, 0)
       {\textbf{By hand}\\[2pt]\scriptsize specs and code by a person; no model};
  \node[stage, fill=teallight, draw=teal]      (b) at (3.5, 0)
       {\textbf{With help}\\[2pt]\scriptsize intent $\to$ spec by P; review by \Hp{} (D1, D2)};
  \node[stage, fill=orangelight, draw=orange]  (c) at (7.0, 0)
       {\textbf{A run}\\[2pt]\scriptsize an agent writes the intents; rounds; asks H};
  \node[stage, fill=corallight, draw=coral]    (d) at (10.5, 0)
       {\textbf{Unsupervised}\\[2pt]\scriptsize \Hp{} answers for H; H reviews after};
  \draw[arrow] (a) -- (b);
  \draw[arrow] (b) -- (c);
  \draw[arrow] (c) -- (d);
  \draw[decorate, decoration={brace, amplitude=6pt, mirror}, draw=teal]
       ($(a.south west)+(0,-0.15)$) -- node[below=7pt, text=teal, font=\scriptsize\itshape]
       {layer 1, with the human as intent author} ($(b.south east)+(0,-0.15)$);
  \draw[decorate, decoration={brace, amplitude=6pt, mirror}, draw=orange]
       ($(c.south west)+(0,-0.15)$) -- node[below=7pt, text=orange, font=\scriptsize\itshape]
       {layer 2, measured by layer 1} ($(d.south east)+(0,-0.15)$);
\end{tikzpicture}
\caption{The spectrum of operation.  The floor runs and blocks at every step.}
\label{fig:spectrum}
\end{figure}

The same engine supports a spectrum of operation (Figure~\ref{fig:spectrum}).
Each step to the right delegates more authorship or more judgement; none of them
lowers the floor.

\subsection{By Hand}
\label{sec:manual}

Without any model, the engine is a gate over specifications and code a person
writes.  New nodes are created as empty skeletons, filled in an editor, and
saved; a HARD change asks before it is saved and is recorded as accepted by the
human.  The gate, drift detection, test evidence, the change set and the earned
status all work, and the pre-commit hook refuses a commit that drifts.  This is
the smallest complete use of the method, and the one that costs nothing.

\subsection{With Help: from Intent to Spec}
\label{sec:assisted}

With a planner endpoint, the human writes what a node is for -- the intent, in
language, with examples -- and P drafts the \texttt{SPEC.md}.  The planner
derives the code globs from the owning container's layout invariants, which are
carried down the node's ownership path, so the build layout is a specification
decision rather than something the coder improvises.  Drafting over an existing
specification regenerates it; revising keeps what the change does not mention.
The draft gate decides who looks at the draft (D0--D2), and every draft shows its
change set.  Every intent is recorded as the node's growth history, with its
author, so that \Hp{} can later judge a changed promise against what the node was
grown for.  The planner is told to say each thing once -- what the architecture
or the ancestors already say reaches the node anyway -- and to keep the prose to
the node as it is now: why a revision changed what it changed is a note recorded
with that revision, not a paragraph added to the specification.  Left alone,
specifications grew to a hundred lines for a component that loads a sound file,
the same prohibition repeated five times; a \emph{compaction} says an existing
one again in fewer words, and the cockpit warns when a promised operation went
missing on the way.  For pending HARD changes, \Hp{} prepares briefs on request, and the
human decides with the brief in front of them; the log records which brief the
human had.

\subsection{An Agent Run}
\label{sec:run}

In a run, the intent author I is an agent.  It reads a manual that tells it how
the graph grows -- iteratively over vertical slices, decomposed as far as it
\emph{knows}, with what it would have to guess left to the slice that needs it
-- and reaches P and C only through the engine's commands: it drafts and revises
through P and implements through C, and writes no specification and no code
itself.  The agents run on the Codex CLI~\cite{codexcli2025}, pointed at the
endpoint configured for their role.

A run goes in \emph{rounds}.  One agent session ends when the model stops --
after a slice, after a failure, or because it believes it is done -- and the
prompt alone (``work until the task is done'') made a run that stopped after one
slice look like a finished one.  After every session, the engine measures the
floor and, on a green floor, \Hp{} checks the goal: it goes through the goal
statement by statement and names \emph{gaps} -- what is missing, never how to
close it -- and, apart from them, what is \emph{waiting}: unmet only because it
rests on a decision only the human may make.  It does not see what the agent
said about its own work, and it must account for the gaps it named in the
previous round before it looks for new ones.  A deterministic rule then decides
(Table~\ref{tab:rounds}); to resume is to continue the same session, with its
memory of the run.

\begin{table}[h]
\centering
\caption{The round rule.  The agent's ``DONE'' is a claim, never the verdict.}
\label{tab:rounds}
\small
\begin{tabularx}{\textwidth}{@{}Xll@{}}
\toprule
\textbf{After a round} & \textbf{Next} & \textbf{Outcome} \\
\midrule
the session failed & stop & error \\
the run itself raised an exception & stop & error (with its reason) \\
a resumed round ran no command & stop & stalled \\
the floor is red & resume & -- (with the failures) \\
floor green, \Hp{} names gaps & resume & -- (with the gaps) \\
floor green, no gap, something waits for the human & stop & asking \\
floor green, no gap, nothing waits & stop & done \\
floor green, \Hp{} unavailable, the agent said DONE & stop & unchecked \\
a resume past the round limit (default 5) & stop & limit \\
\midrule
the agent said BLOCKED: measured all the same -- green, no gap, and something
waits: \emph{asking}; otherwise & stop & blocked \\
\bottomrule
\end{tabularx}
\end{table}

Two distinctions in this rule were learned from runs.  \emph{Waiting for the
human is not being blocked}: a run that reported itself blocked over a green
floor, with six requests and waivers open, was in fact only waiting -- and the
record should say so.  And \emph{what waits for the human is not a gap}: sending the agent back
for something only the human may decide -- in one run, a root node it was not
allowed to change -- only burns rounds.

When a run ends, the engine records the tests of the final state, lets \Hp{}
decide what the delegation allows, and has \Hp{} brief every pending HARD
change, so that the human finds the homework done.  Everything a run did is
recorded in the project: the event stream, a readable log, and which model at
which endpoint played which role.  A run that crashes ends all the same, as
\emph{error}, with the reason, where it was raised and the traceback in that
record -- not only on a terminal nobody may be watching.  A run that ends
\emph{asking} leaves its work
in place; the human answers the questions, and the next run continues on the
same graph.

The last word of a round is also the human's to ask for without a run: the
\emph{goal check} runs the engine's floor and, on a green one, \Hp{} against the
goal -- no agent, no session, nothing changed but the recorded test evidence.
It ends \emph{done}, \emph{gaps} (with the list, each with where it sits: \Hp{}
quotes a few words from the specification or the goal that shows it, and the
engine finds them to the line, so that the human can open the place),
\emph{asking}, \emph{red} or \emph{unchecked}; \emph{asking} counts everything that waits for the human, a
pending HARD change too, so that \emph{done} means nothing does.  Its use is the
moment after the human worked through the requests, or changed something by
hand.

\paragraph{Asked after, not before.}  Where the human is present, the human is
asked first: in the cockpit, saving asks before a HARD change is written, and
the draft gate reviews a draft before it is saved -- all of it before any code.
A run is unattended by definition, so it cannot ask at the moment a question
comes up; it goes on credit instead.  A HARD change stays pending against the
accepted state and keeps its nodes from \emph{verified}; a decision request is
recorded, and the work rests on its proposal.  The human decides at the end
(Table~\ref{tab:when}).

\begin{table}[h]
\centering
\caption{Asking before the code, or after it.  A run asks after.}
\label{tab:when}
\small
\begin{tabularx}{\textwidth}{@{}l>{\raggedright\arraybackslash}X>{\raggedright\arraybackslash}X@{}}
\toprule
& \textbf{Asked before the code} & \textbf{Asked after it (a run)} \\
\midrule
\textbf{The human approves} & The human waited; the run stood still. & Nothing is
  lost. \\
\textbf{The human rejects} & No code was built in vain. & The work built on it is
  taken back -- the gate holds it until then, and the change set shows what. \\
\textbf{The human has to be} & There, during the run. & There at the end. \\
\bottomrule
\end{tabularx}
\end{table}

We considered, and did not build, a run that asks \emph{early}: the coder
refusing to implement a node whose specification rests on an open request or an
unaccepted HARD change, so that the run ends \emph{asking} before the code.  It
would make the human the clock of the run -- which is what working in the
cockpit already is.  And the expensive case does not need it: a decision that
would invalidate much work is one the human can see coming, and what the human
wants decided before any code is decided before the run, in the goal or as an
invariant of the architecture.  What is still open when a run starts is what
nobody foresaw, and asking about it afterwards costs, at worst, the work built on
one proposal.  The unsupervised run is the third way: \Hp{} answers between
rounds, and the human reviews afterwards.

\subsection{Unsupervised}
\label{sec:unsupervised}

A supervised run stops at the first thing only the human may decide that nothing
else can go on without, and in practice that is early: three of the later runs
ended with two, nine and six items waiting for the human.  In an \emph{unsupervised}
run, \Hp{} stands in for the human between rounds:

\begin{itemize}[leftmargin=*]
  \item \textbf{Requests.}  \Hp{} answers every open decision request -- it
        accepts or rejects it \emph{as worded}, and never rewords it.  The
        answer is recorded as an ADR in \texttt{ARCHITECTURE.md} that names
        \Hp{} as the one who gave it, and the request is removed.  An answer
        \Hp{} is unsure of (confidence below~0.6) is not given; that request
        stays open for the human.
  \item \textbf{Delegated changes.}  \Hp{} decides the pending HARD changes
        inside the delegation, as in a supervised run -- which is why an
        unsupervised run requires the delegation to be on.
  \item \textbf{Rework.}  A rejection sends the agent back, since its work
        assumed the answer; so does an acceptance that targets a node, since the
        ADR authorises a change the agent now has to make -- the one way an
        unsupervised run may change a node it was given.
\end{itemize}

\begin{solutionbox}
\textbf{Nothing new keeps it the human's.}  An answer by \Hp{} is a line in
\texttt{ARCHITECTURE.md}; a change to that file is HARD and outside the
delegation.  So every answer waits in the human's pending list until the human
accepts it -- or rejects it, which blocks the gate until the work is taken back.
What stays the human's even here: the goal, the zones the human declared, and
every answer, afterwards.
\end{solutionbox}

Answering a request is authoring intent above a node.  The implementation
records this as the second, bounded exception to the rule that agents author no
intent -- the first being the node-level intents an agent writes in any run --
and makes unsupervised a setting the human chooses per run.

\subsection{Setting Up a Project}
\label{sec:setup}

A project may live in any directory.  The New Project dialog, or its
command-line equivalent, lays the floor there and records the answers it was
given, so that the same floor can be laid again elsewhere: a second attempt at
a task is a new directory laid like the first, not the first one reset.  The
agent manuals and the goal are rendered from templates that belong to the
method, and can be rendered again when the templates change.

Every role that calls a model is configured the same way, by its own lines in
one file: an OpenAI-compatible base URL, a key and a model -- and for the two
agents, I and C, the harness that runs them; optionally, how hard the model
reasons.  Nothing falls back: not one role to another, not a role to a provider
built into the code, not an agent to the harness's own user configuration, which
the engine does not load -- so the file is the whole configuration, the same on
every machine.  Two roles on the
same model repeat the same values, which makes it visible that they are the
same, and the cockpit says so when \Hp{} reviews what its own model wrote.
Whatever keeps a role from being called -- an unset value, a base URL that is
not a URL, a key that is still an example's placeholder, the harness missing
from the path -- is reported before anything starts.

\section{Example: A Number-Guessing Game}
\label{sec:example}

We follow one small project from an empty directory to a goal the engine and \Hp{}
call met -- in the cockpit, by hand and with help, the first half of the spectrum
of Section~\ref{sec:operation} -- and then give the same goal to two agent runs,
the second half.  Every number below is from those sessions; the cockpit's manual
walks through the first with a screenshot of every step.
All roles that call a model ran on the same hosted model with a high reasoning
effort, so \Hp{} reviewed what its own model had written -- a limitation the
cockpit states with every review, and one this example shares.

\paragraph{The floor.}  The program: the computer picks a secret number from 1 to
100, the player types guesses, and the game answers \texttt{higher}, \texttt{lower}
or \texttt{correct -{}- N guesses}.  The New Project dialog laid the floor -- a root,
one container \texttt{game} with a Python layout and its build manifest, and six
invariants: the four every project starts with and two the human's answers added
(nothing is kept between runs; the program never touches the network).  The goal
was typed as three behaviours, two statements that must always hold and one
exclusion; the planner proposed a fuller goal -- twelve behaviours, ten
statements, eight exclusions -- which the human kept.  That \texttt{TASK.md} is
the yardstick of everything that follows.

\paragraph{Two components, drafted and coded.}  The human created two
components and wrote an intent for each, with examples: \texttt{game.rules}, the
game without a terminal, and \texttt{game.cli}, the command line around it.  P
drafted both specifications; at draft gate D1, \Hp{} approved the first with
confidence 0.72 and five advisory findings, and the human saved both.  The change
set classified each as a new component -- SOFT -- and named the next step,
\emph{Codex game.rules, then Run Tests}.  For the command line, P declared by
itself that it depends on the rules; the drift check later found the matching
import.  C implemented \texttt{game.rules} in under two minutes with 17 tests,
then \texttt{game.cli}.  The whole build passed 43 tests, and the gate passed:
seven owned source files, four import edges, no error, no warning.

\begin{figure}[h]
\centering
\includegraphics[width=\textwidth]{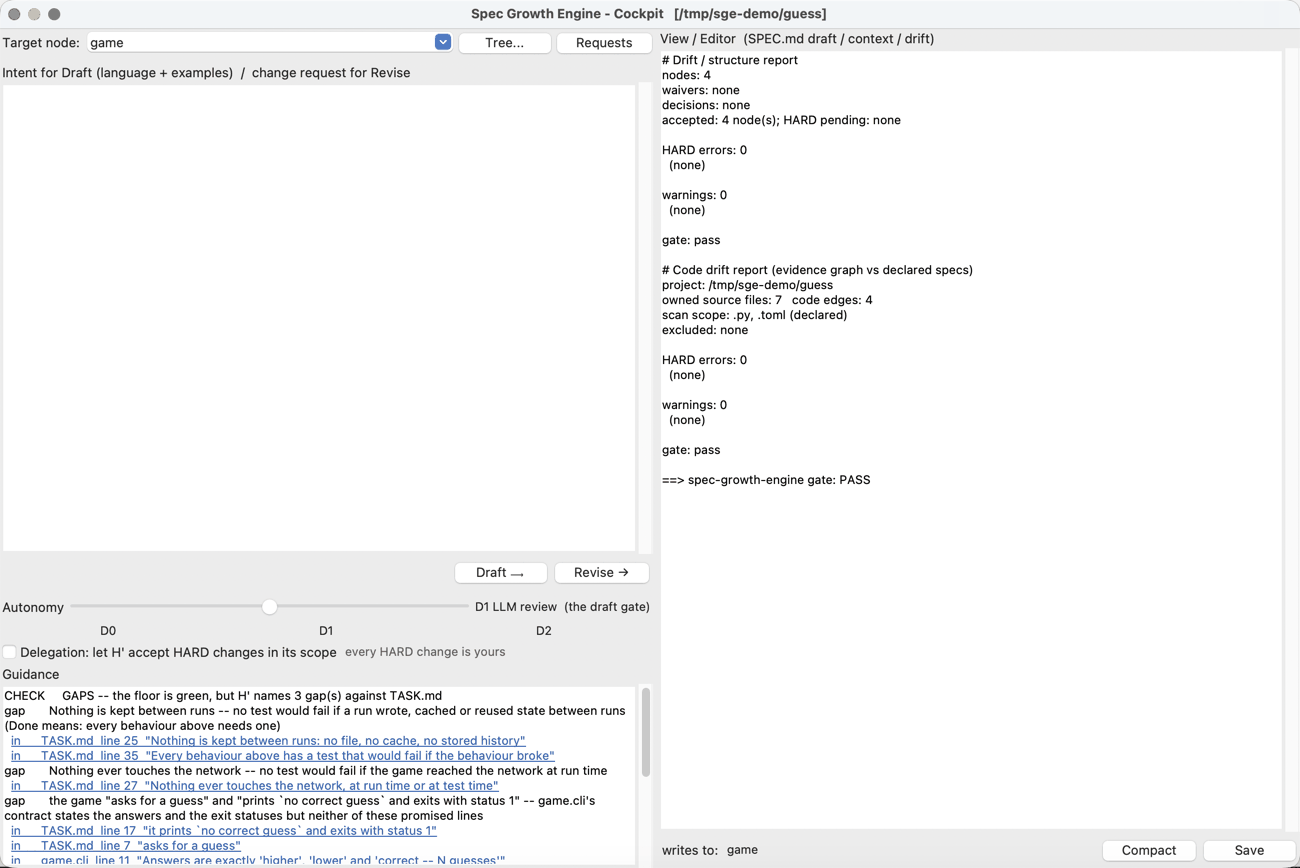}
\caption{The cockpit after the first goal check: three gaps in the guidance pane,
each with the place that shows it; a click opens the place, the words marked.}
\label{fig:check}
\end{figure}

\paragraph{A green floor is not a met goal.}  The goal check -- the engine's
floor, then \Hp{} against \texttt{TASK.md}, with no agent -- returned three gaps
(Figure~\ref{fig:check}):

\begin{lstlisting}
CHECK  GAPS -- the floor is green, but H' names 3 gap(s) against TASK.md
gap    Nothing is kept between runs -- no test would fail if a run wrote,
       cached or reused state between runs
  in   TASK.md  line 25  "Nothing is kept between runs: no file, no cache, ..."
gap    Nothing ever touches the network -- no test would fail if the game
       reached the network at run time
  in   TASK.md  line 27  "Nothing ever touches the network, at run time ..."
gap    the game "asks for a guess" and "prints `no correct guess` and exits
       with status 1" -- game.cli's contract states neither
  in   TASK.md  line 17  "it prints `no correct guess` and exits with status 1"
  in   game.cli  line 11  "Answers are exactly 'higher', 'lower' and ..."
\end{lstlisting}

Two of them are statements that must always hold and that no test would catch
breaking -- what \emph{Done means} asks for and a passing build cannot show; the
third is behaviour the goal promises and no contract states.  Each gap names
where it sits, found to the line, and the human opens the place with a click
instead of searching for it.

\paragraph{Closing the gaps.}  One revision of \texttt{game.cli} closed all
three: a change request in the human's words, which P applied to the saved
specification while keeping everything else.  P's note -- shown with the change
set and recorded with the intent, not written into the specification -- called
it additive, and the change set agreed: two promises and three acceptance
criteria added, SOFT.  The human saved it, and the coder ran on the node and on
whoever uses it, here no one.  C found the code printing a different prompt than
the revised contract now promised, corrected code and tests, and the whole build
passed 46 tests and the gate.

\paragraph{The check reads the whole goal again.}  The second goal check found
the three gaps closed and named a new, smaller one: without a seed, nothing
showed that the secret really varies -- a fixed default would have passed every
test.  A second revision added that promise, and P's note said what else it had
to change: an existing criterion, identical output for identical runs, was
narrowed to runs with the same explicit seed.  The coder ran again, and the third
check returned \texttt{DONE} -- the tests and the gate pass, \Hp{} finds no gap,
nothing waits for the human -- over 47 tests (Table~\ref{tab:example}).

\begin{table}[h]
\centering
\caption{The example's three goal checks.}
\label{tab:example}
\small
\begin{tabular}{@{}lccl@{}}
\toprule
\textbf{After} & \textbf{Tests} & \textbf{Gate} & \textbf{\Hp{}} \\
\midrule
the first code & 43 & pass & 3 gaps \\
the first revision of \texttt{game.cli} & 46 & pass & 1 gap \\
the second revision of \texttt{game.cli} & 47 & pass & done \\
\bottomrule
\end{tabular}
\end{table}

\paragraph{A compaction is a breaking change.}  Specifications revised several
times grow, and the cockpit can say one again in fewer words.  Compacting
\texttt{game.rules} took it from 55 to 50 lines and kept every promised operation
-- and the change set called it HARD: reworded promises that \texttt{game.cli}
uses count as broken, because the engine reads a contract as text and cannot
tell a rewording from a change.  That is the gate failing safe; the human reads
the before-and-after pairs and decides.  In the example it was not saved.

\paragraph{What the example shows.}  A floor that is green -- tests, gate, an
earned status -- is necessary and not sufficient: the goal check found three
unmet statements over 43 passing tests.  The specification, not the goal, is what
the coder builds from: every gap was closed in a contract, and the code followed.
The reviewer converges rather than settles at once: each check reads the whole
goal again and may find something new, and here two rounds were enough.  And the
engine's caution is visible rather than silent: a compaction that changed nothing
in substance was still a breaking change, shown entry by entry.

\paragraph{The same goal, given to two runs.}  Two more projects were laid with
New Project Like -- the floor above and its \texttt{TASK.md}, nothing else -- and
each was given to an agent run: one supervised, one unsupervised with the
delegation on.  Both ended after one round, the intent author having claimed
\texttt{DONE} over a green floor; neither ended \texttt{DONE}
(Table~\ref{tab:runs}).

\begin{table}[h]
\centering
\caption{The same goal three times: by hand with help, and in two agent runs.}
\label{tab:runs}
\small
\begin{tabular}{@{}lccc@{}}
\toprule
 & \textbf{By hand} & \textbf{Supervised run} & \textbf{Unsupervised run} \\
\midrule
Rounds & -- & 1 & 1 \\
Wall time & -- & 15 min & 32 min \\
Components & 2 & 3 & 4 \\
Tests & 47 & 130 & 229 \\
Gate & pass & pass & pass \\
Gaps named by \Hp{} & none (third check) & none & none \\
Ended & done & asking & unchecked \\
Waits for the human & nothing & a decision request & a decision made \\
 & & & in the human's place \\
\bottomrule
\end{tabular}
\end{table}

\paragraph{Supervised.}  The intent author grew three components --
\texttt{game.secret}, the only place chance enters; \texttt{game.rules};
\texttt{game.cli} -- each drafted through P and coded by C, and had C write the
container's console script.  Before claiming \texttt{DONE} it drove the real
program through every behaviour of \texttt{TASK.md}.  The run stopped
\texttt{ASKING}: \texttt{game.cli} had filed a decision request on the container
-- record in \texttt{game}'s contract that the console script \texttt{guess}
starts \texttt{game.cli} -- and the container is the human's.  \Hp{} named no gap
and put one behaviour of the goal on hold, the program starting with
\texttt{uv run -{}-project game guess}: it works and a test pins it, but it rests on
the open request.  One round, one question for the human.

\paragraph{Unsupervised.}  The second run cut the game into four components: a
\texttt{game.parsing} shared by the seed and the guess, so that the two checks
cannot drift apart, then \texttt{game.secret}, \texttt{game.play} and
\texttt{game.cli}.  It corrected P's drafts as it went -- filled an empty
contract, replaced hand-rolled arithmetic with a seeded generator, added a test
for ``nothing is kept between runs'' -- and, driving the program through pipes one
line at a time, found a defect no test had caught: standard output, block-buffered
on a pipe, deadlocked a step-by-step driver.  It added the promise to flush every
line to \texttt{game.cli}'s contract -- SOFT, cleared by E -- and C added two tests
that fail without it.

One planner slip cost the most time.  P had nested a design note inside
\texttt{game.play}'s contract, and repeated revisions did not remove it: a
revision returns every key the planner leaves out -- omission is not deletion, the
guard against a planner that drops a promise unasked -- and so it cannot delete
one on request either.  The agent overwrote the value with a pointer instead.
That changed a contract \texttt{game.cli} uses: HARD.  With the delegation on,
\Hp{} wrote a brief -- accept, confidence 0.70 -- E found the one consumer
migrated, and \Hp{} accepted the change in the human's place.

The run then ended \texttt{UNCHECKED}: the floor was green, but \Hp{}'s goal check
was cut off at the provider's default output limit, and the run said so instead
of calling the goal met.  Repeated afterwards with Check Goal, the check named no
gap and returned \texttt{ASKING}: the acceptance \Hp{} had made in the human's
place waits for the human's review.  The run's closing line had listed no such
decision -- it records what \Hp{} decides \emph{between} rounds, and this one was
made inside the round, at the agent's request; the growth record held it, and the
goal check put it in front of the human.

\paragraph{What the runs show.}  The same goal, floor and model grew three
graphs -- two, three and four components: the goal fixes what the program does,
not how it is cut.  Both intent authors drove the real program through
\texttt{TASK.md} before claiming \texttt{DONE}, and neither left a gap for \Hp{}
to name -- the work the reviewer's three checks did in the walkthrough.  Neither
run ended \texttt{DONE}, and neither should have: each stopped at what only the
human may close -- a decision on the human's container, and the review of a
decision made in the human's place.  Unsupervised moves the human from before to
after; it does not remove them.  And the runs found weak spots in the engine
itself -- a revision that cannot delete, a goal check exposed to a provider's
default output limit, a closing line that missed a decision made inside a round
-- each surfacing as a stated finding or a cautious outcome, \texttt{UNCHECKED} or
\texttt{ASKING}, never as a silent \texttt{DONE}.  All three were fixed after these
runs: a revision deletes a key the planner names on purpose, P and \Hp{} may each
set their own output limit, and the closing line reads what \Hp{} decided inside
a round from the audit log.

\paragraph{Availability.}
The engine, the command-line interface, the cockpit with its manual and the
method documents are available under the Apache License~2.0 at\\
\url{https://github.com/grabow/spec-growth-engine}.

\bibliographystyle{plainnat}
\bibliography{refs}

\end{document}